\documentclass[aps,prl,twocolumn,showpacs,superscriptaddress]{revtex4}

\usepackage{graphicx}

\begin{document}

\title{Spin Waves and Spatially Anisotropic Exchange Interactions in the $\mathrm{S=2}$ Stripe Antiferromagnet Rb$_{0.8}$Fe$_{1.5}$S$_2$}
\author{Meng Wang}
\email{wangm@berkeley.edu}
\affiliation{Department of Physics, University of California, Berkeley, California 94720, USA }
\author{P. Valdivia}
\affiliation{Department of Physics, University of California, Berkeley, California 94720, USA }
\author{Ming Yi}
\affiliation{Department of Physics, University of California, Berkeley, California 94720, USA }
\author{J. X. Chen}
\affiliation{School of Physics and Engineering, Sun Yat-Sen University, Guangzhou 510275, China }
\author{W. L. Zhang}
\affiliation{Institute of Physics, Chinese Academy of Sciences, P. O. Box 603, Beijing, 100190, China }
\author{R. A. Ewings}
\affiliation{ISIS Facility, STFC Rutherford Appleton Laboratory, Chilton, Didcot OX110QX, United Kingdom }
\author{T. G. Perring}
\affiliation{ISIS Facility, STFC Rutherford Appleton Laboratory, Chilton, Didcot OX110QX, United Kingdom }
\author{Yang Zhao}
\affiliation{NIST Center for Neutron Research, National Institute of Standards and Technology, Gaithersburg, Maryland 20899, USA }
\affiliation{Department of Materials Science and Engineering, University of Maryland, College Park, Maryland 20742, USA }
\author{L. W. Harriger }
\affiliation{NIST Center for Neutron Research, National Institute of Standards and Technology, Gaithersburg, Maryland 20899, USA }
\author{J. W. Lynn}
\affiliation{NIST Center for Neutron Research, National Institute of Standards and Technology, Gaithersburg, Maryland 20899, USA }
\author{E. Bourret-Courchesne}
\affiliation{Materials Science Division, Lawrence Berkeley National Laboratory, Berkeley, California 94720, USA }
\author{Pengcheng Dai}
\affiliation{Department of Physics and Astronomy, Rice University, Houston, Texas 77005, USA}
\author{D. H. Lee}
\affiliation{Department of Physics, University of California, Berkeley, California 94720, USA }
\affiliation{Materials Science Division, Lawrence Berkeley National Laboratory, Berkeley, California 94720, USA }
\author{D. X. Yao}
\affiliation{School of Physics and Engineering, Sun Yat-Sen University, Guangzhou 510275, China }
\author{R. J. Birgeneau}
\affiliation{Department of Physics, University of California, Berkeley, California 94720, USA }
\affiliation{Materials Science Division, Lawrence Berkeley National Laboratory, Berkeley, California 94720, USA }
\affiliation{Department of Materials Science and Engineering, University of California, Berkeley, California 94720, USA }

\begin{abstract}
An inelastic neutron scattering study of the spin waves corresponding to the stripe antiferromagnetic order in insulating Rb$_{0.8}$Fe$_{1.5}$S$_2$ throughout the Brillouin zone is reported.  The spin wave spectra are well described by a Heisenberg Hamiltonian with anisotropic in-plane exchange interactions. Integrating the ordered moment and the spin fluctuations results in a total moment squared of $27.6\pm4.2\mu_B^2$/Fe, consistent with $\mathrm{S \approx 2}$. Unlike $X$Fe$_2$As$_2$ ($X=$ Ca, Sr, and Ba), where the itinerant electrons have a significant contribution, our data suggest that this stripe antiferromagnetically ordered phase in Rb$_{0.8}$Fe$_{1.5}$S$_2$ is a Mott-like insulator with fully localized $3d$ electrons and a high-spin ground state configuration. Nevertheless, the anisotropic exchange couplings appear to be universal in the stripe phase of Fe pnictides and chalcogenides.
\end{abstract}

\maketitle




Superconductivity emerges in the vicinity of antiferromagnetism (AFM) in both copper based and iron based high-transition temperature (high-$T_c$) superconductors\cite{kastner, birgeneau, davis}. The AFM in these systems share several similarities: antiferromagnetic order in layered parent compounds, a spin resonance mode in the superconducting state, and the presence of spin fluctuations throughout the doping-temperature phase diagrams\cite{pengcheng}. However, the AFM in the cuprate high-$T_c$ and iron-based superconductors could have different origins. The parent compound of the copper oxide superconductors is a Mott insulator with $\mathrm{S=1/2}$ local moments\cite{hayden}. In the iron pnictides the parent compounds are bad metals with several bands crossing the Fermi energy. The stripe AF ordering wavevectors coincide with the wave vectors connecting the centers of the electron and hole Fermi surfaces\cite{dongjing}. In fact many view the AF order as due to the Fermi surface nesting.

From a localized point of view, with 6 electrons in the iron $3d$ orbitals of Fe$^{2+}$, the maximum total spin is $\mathrm{S=2}$. This spin state can be realized when the Hund's Rule coupling energy, $J_H$, dominates over the crystal-field splitting associated with the Fe$M_4$ ($M=$ pnictigens or chalcogens) structural unit. On the other hand, a crystal field splitting $\Delta_{\mathrm{CF}}$ comparable to the Hund's coupling $J_H$ can lead to an intermediate-spin $\mathrm{S=1}$ state. In the large crystal field extreme, the $3d^6$ ions of Fe$^{2+}$ will form a low-spin singlet $\mathrm{S=0}$ state\cite{qimiao08, frank, haule}. In the presence of itinerant carriers the spin must be less than $\mathrm{S=2}$ due to charge fluctuations. Thus, while the observation of an intermediate-spin state $\mathrm{S=1}$ does not rule out the presence of itinerant carriers, the observation of $\mathrm{S=2}$ would require the system to be predominately localized. Not surprisingly, the various values of ordered moments observed in different iron-based materials have been interpreted in terms of both the local moment picture and the itinerant carrier picture\cite{qimiao08, daniel, haule, frank}. As to the value of the fluctuating local moment, inelastic neutron scattering experiments combined with the moment sum rule revealed an increase of $S$ from $\mathrm{S\approx1}$ at 10 K to $\mathrm{S\approx3/2}$ at 300 K for Fe$_{1.1}$Te and a constant $\mathrm{S=1/2}$ for BaFe$_2$As$_2$\cite{igor, mengshu, leland}. In addition, an $X$-ray emission spectroscopy study was interpreted to imply that the iron spin-state varied between $\mathrm{S=0}$ and 2 in the rare-earth doped Ca$_{1-x}$RE$_x$Fe$_2$As$_2$ as a function of temperature\cite{yjkim}. In contrast, studies of the spin wave excitations in $X$Fe$_2$As$_2$  ($X=$ Ca, Sr, and Ba) have found significant special weight contributions from both itinerant carriers and local moments\cite{mcqueeney2, junnaturephys, leland, russell}. These findings suggest that the magnetism of the iron pnictides and chalcogenides should be understood from a point of view where both itinerant carriers and local moments coexist.

The substitution of sulfur for selenium progressively suppresses the superconductivity in K$_{0.8}$Fe$_{y}$Se$_{2-z}$S$_{z}$ and decreasing the Fe content results in an insulating ground state\cite{lei,jgguo}. We found a stripe AF order in insulating Rb$_{0.8}$Fe$_{1.5}$S$_{2}$ with a rhombic iron vacancy order and a strikingly similar N$\acute{\mathrm{e}}$el temperature of $T_N=275$ K with a moment size of $M=2.8\pm0.5\mu_B$ as that in K$_{0.81}$Fe$_{1.58}$Se$_{2}$\cite{meng, jzhao}. Photoemission measurements revealed a 980 meV charge gap in the rhombic iron vacancy ordered phase\cite{fchen}. The stripe AF order was proposed as a candidate parent compound for the superconducting phase in $A_{0.8}$Fe$_{y}$Se$_{2}$ ($A=$ alkali metal)\cite{jzhao,si}. It is therefore important to characterize the spin waves associated with the stripe AF order in insulating Rb$_{0.8}$Fe$_{1.5}$S$_{2}$ in order to understand the nature of its magnetism.

\begin{figure}[t]
\includegraphics[scale=0.43]{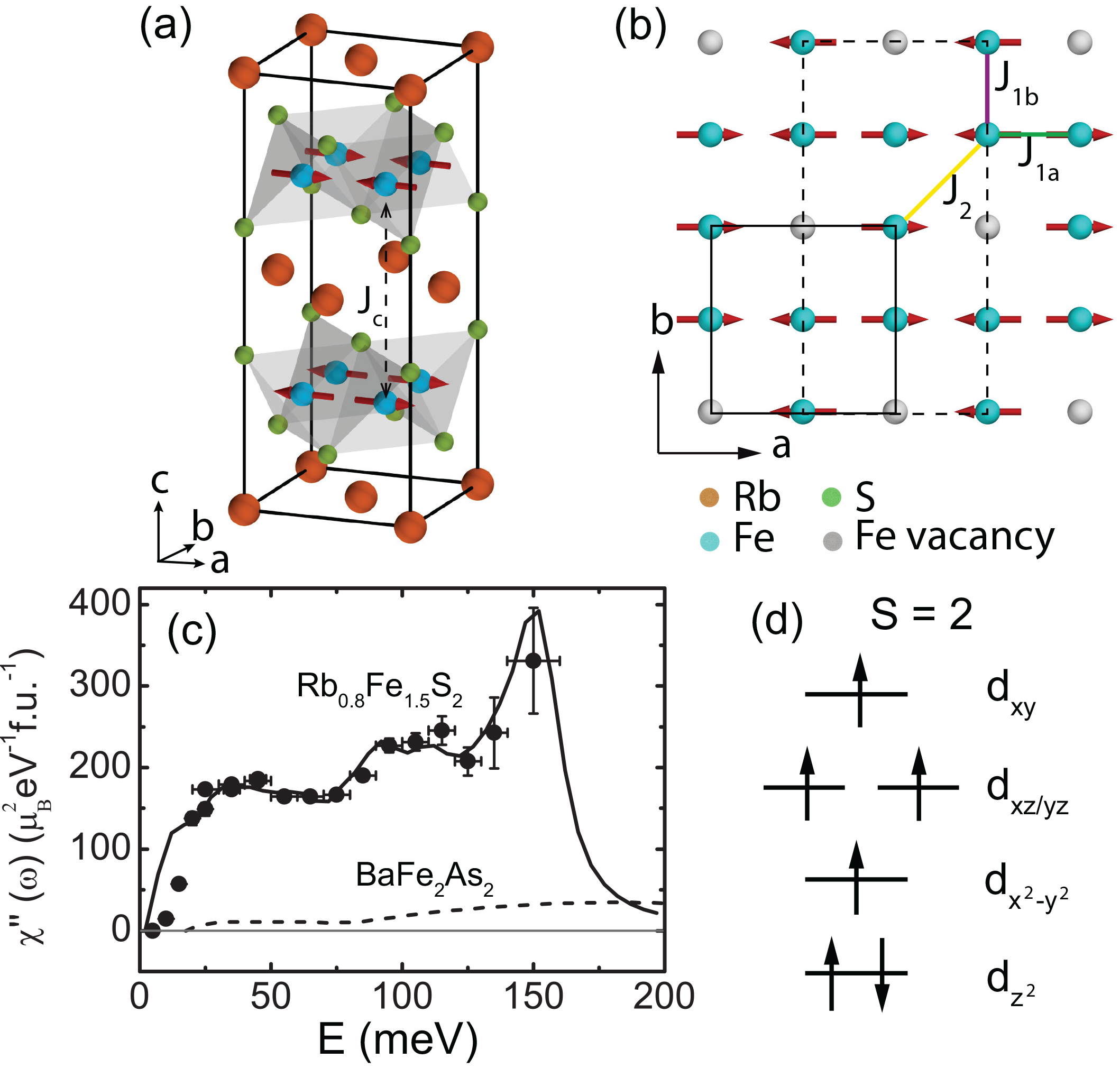}
\caption{ (color online). Three- (a) and Two- (b) dimensional structures of the stripe AF order with rhombic iron vacancy order in Rb$_{0.8}$Fe$_{1.5}$S$_{2}$. We use the orthorhombic unit cell as shown by the solid square in (b) with lattice parameters of $a=5.58$ \AA, $b=5.39$ \AA, and $c=13.889$ \AA. The wave vector $Q$ is defined as $Q=[H, K, L]=(2\pi H/a, 2\pi K/b, 2\pi L/c)$ in reciprocal lattice units (r.l.u). The dashed rectangle is the real magnetic unit cell. (c) Dynamic susceptibility $\chi^{\prime\prime}(\omega)$ as a function of energy with $E_i=35$, 170, and 250 meV at 8 K. The solid line is computed by the model discussed in the text. The dashed line is the dynamic susceptibility of BaFe$_2$As$_2$ from Ref. \cite{mengshu}. (d) A candidate for the high-spin ground state configuration of the stripe AF order\cite{qimiao08}. }
\label{fig1}
\end{figure}

In this paper, we report inelastic neutron scattering studies of the spin wave excitations of the stripe AF order in insulating Rb$_{0.8}$Fe$_{1.5}$S$_{2}$. Only the spin excitations associated with the stripe AF order are observed in our experiment, suggesting a nearly 100$\%$ stripe AF order volume fraction. In the presence of iron vacancy order, there are six iron atoms per magnetic unit cell. Hence, one expects three doubly-degenerate spin wave branches. The first acoustic and the second optical branches are observed clearly in both momentum and energy scans in our experiment which uses $E_i\leq250$ meV. The third branch is flat in momentum space and can only be observed by scans in energy. By fitting the spin excitation spectrum to a Heisenberg Hamiltonian with spatially anisotropic exchange couplings ($SJ_{1a}=42\pm5$, $SJ_{1b}=-20\pm2$, $SJ_2=17\pm2$, $SJ_c=0.29\pm0.05$ and $SJ_s=0.09\pm0.02$ meV), all of the branches of the spin excitations can be accurately described. Furthermore, the total dynamic spin fluctuation moment spectrum is calculated to be $\langle m\rangle^2\approx20\mu_B^2/$Fe, similar to that in the block insulating AF Rb$_{0.89}$Fe$_{1.58}$Se$_{2}$\cite{miaoyin}. After including the contribution of the ordered moment $2.8\mu_B$, we estimate the spin to be $\mathrm{S=2}$. Knowing the stripe AF order is an insulator with a large charge gap ($\sim1$ eV), the spin $\mathrm{S=2}$ suggests that all Fe $3d$ electrons are fully localized. 

\begin{figure*}[t]
\includegraphics[scale=0.55]{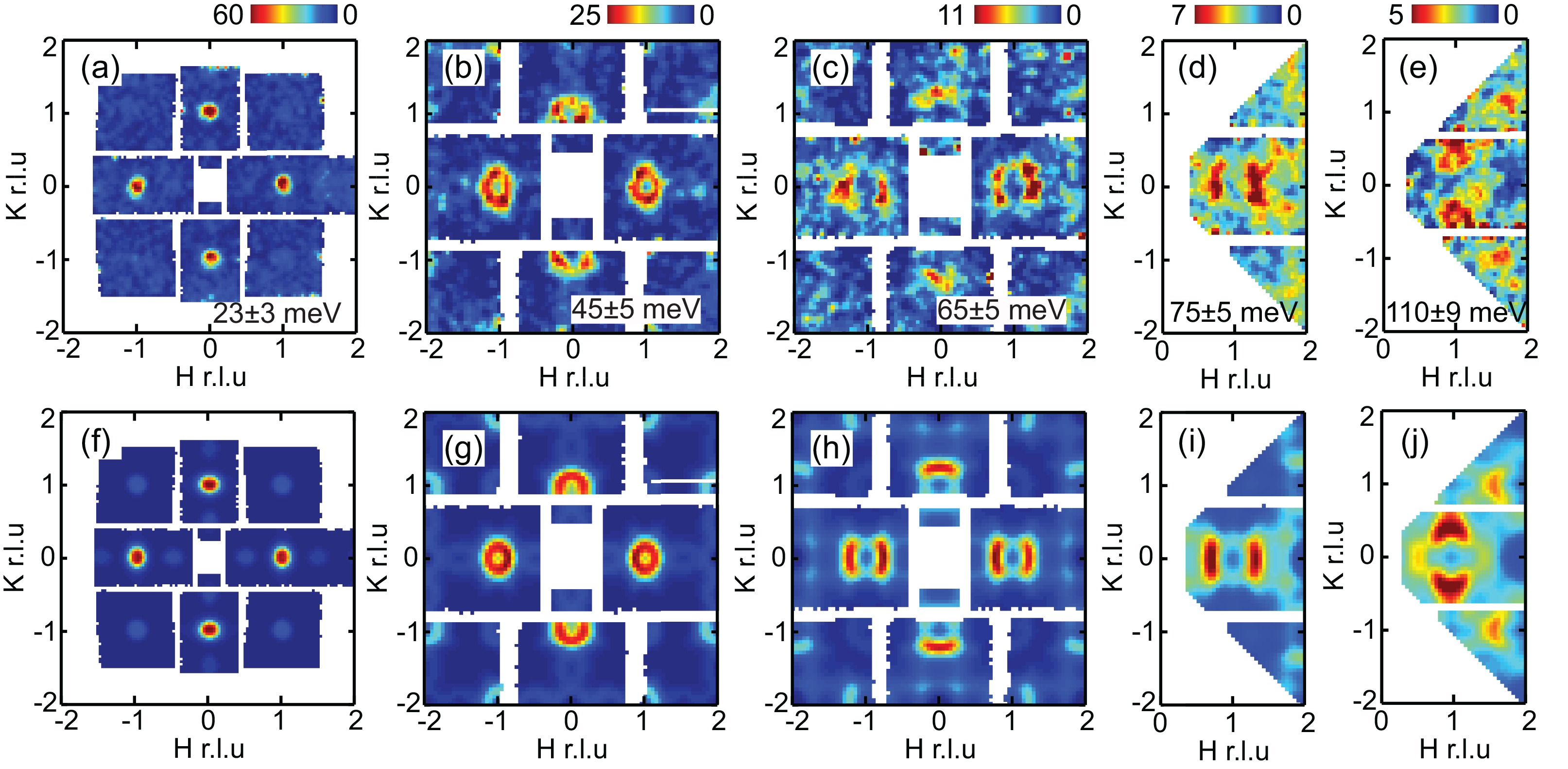}
\caption{ (color online). Constant energy slices in the $[H, K]$ plane of the spin waves at energies of (a) $E=23\pm3$ meV with $E_i=80$ meV, and (b) $E=45\pm5$, (c) $E=65\pm5$, (d) $E=75\pm5$, (e) $E=110\pm9$ meV with $E_i=250$ meV, all at 8 K. (f-j) Simulations of spin excitations at the identical energies as in (a-e) using the exchange couplings from the best fits to the experimental data. The simulations were convoluted with the instrumental resolution. The color bar is the same for each energy transfer in units of $mbarSr^{-1}meV^{-1}f.u.^{-1}$. }
\label{fig2}
\end{figure*}

Our experiments were carried out on the MAPS time-of-flight (TOF) chopper spectrometer at the Rutherford-Appleton Laboratory, Didcot, UK, and the BT-7 thermal triple-axis spectrometer at the NIST Center for Neutron Research, Gaithersburg, USA. We coaligned 1.5 grams of single crystals with a mosaic of 1.5$^{\circ}$ full width at half maximum for the two experiments. The stripe AF order with $M=2.8\pm0.5\mu_B$ moments and rhombic iron vacancy order  has been reported elsewhere\cite{meng}. For the TOF experiment at MAPS, we aligned the $c$ axis of the sample parallel to the incident beam at energies of $E_i=35, 80, 170$ and 250 meV at 8 K. The intensities were normalized to absolute units by vanadium incoherent scattering. For the low energy neutron scattering measurements performed at BT-7, we fixed the final energy at 14.7 meV, with horizontal collimations of open-$80^\prime$-$S$-$80^\prime$-$120^\prime$, where $S=$ sample, and two pyrolytic graphite filters after the sample\cite{bt7}. 

We show spin excitations in the $[H, K]$ plane at various energies in Fig.\ref{fig2} (a-e). The spin excitations stem from the AF wave vectors, disperse outwards and separate into two arcs at $E=65\pm5$ and $75\pm5$ meV. At the energy of $110\pm9$ meV, the wave vectors rotate $90^{\circ}$. Weak spin excitations at $Q=(\pm1\pm0.5, 0)$,  ($0, \pm1\pm0.5$) and ($\pm1, \pm1$) in Fig. \ref{fig2} (a) can also be observed.

To describe the spin waves in Rb$_{0.8}$Fe$_{1.5}$S$_{2}$, we employed a Heisenberg model with in-plane nearest-($J_{1a}$, $J_{1b}$), and next-nearest-($J_2$) neighbor exchange couplings, together with the coupling between layers, $J_c$, as shown in Fig. \ref{fig1} (a) and Fig. \ref{fig1} (b), and the single ion anisotropy term, $J_s$. The Hamiltonian can be written as:
 \begin{equation} 
  \hat{H}=\frac{J_{r,r^\prime}}{2}\sum_{r,r^\prime}\bf{S}_r\cdot S_{r^\prime}-\it{J_s}\sum_r(\bf{S}_{r}^z)^2,
  \label{eq1}
 \end{equation}
where $J_{r,r^\prime}$ are the effective exchange couplings and $(r, r^\prime)$ label the iron sites\cite{daoxin}. The spin wave excitation spectrum can be expressed analytically by solving Eq. (1) using the linear spin wave approximation\cite{junnaturephys, leland,russell,miaoyin}. We fit the data and convolute the instrumental resolution using the Tobyfit program\cite{toby}. From the best fit to the experimental data,  we determine the parameters as $SJ_{1a}=42\pm5$, $SJ_{1b}=-20\pm2$, and $SJ_2=17\pm2$ meV, and for computational convenience an energy independent damping $\Gamma=7\pm2$ meV. The widths of the spin wave peaks in $H$ and $K$ were close to being instrumental resolution limited as expected for an insulator; this also holds true for K$_{0.81}$Fe$_{1.58}$Se$_2$\cite{bob, jun2}. The simulations with the fit parameters at the identical energies of Fig. \ref{fig2} (a-e) are presented in Fig. \ref{fig2} (f-j).

\begin{figure}[t]
\includegraphics[scale=0.42]{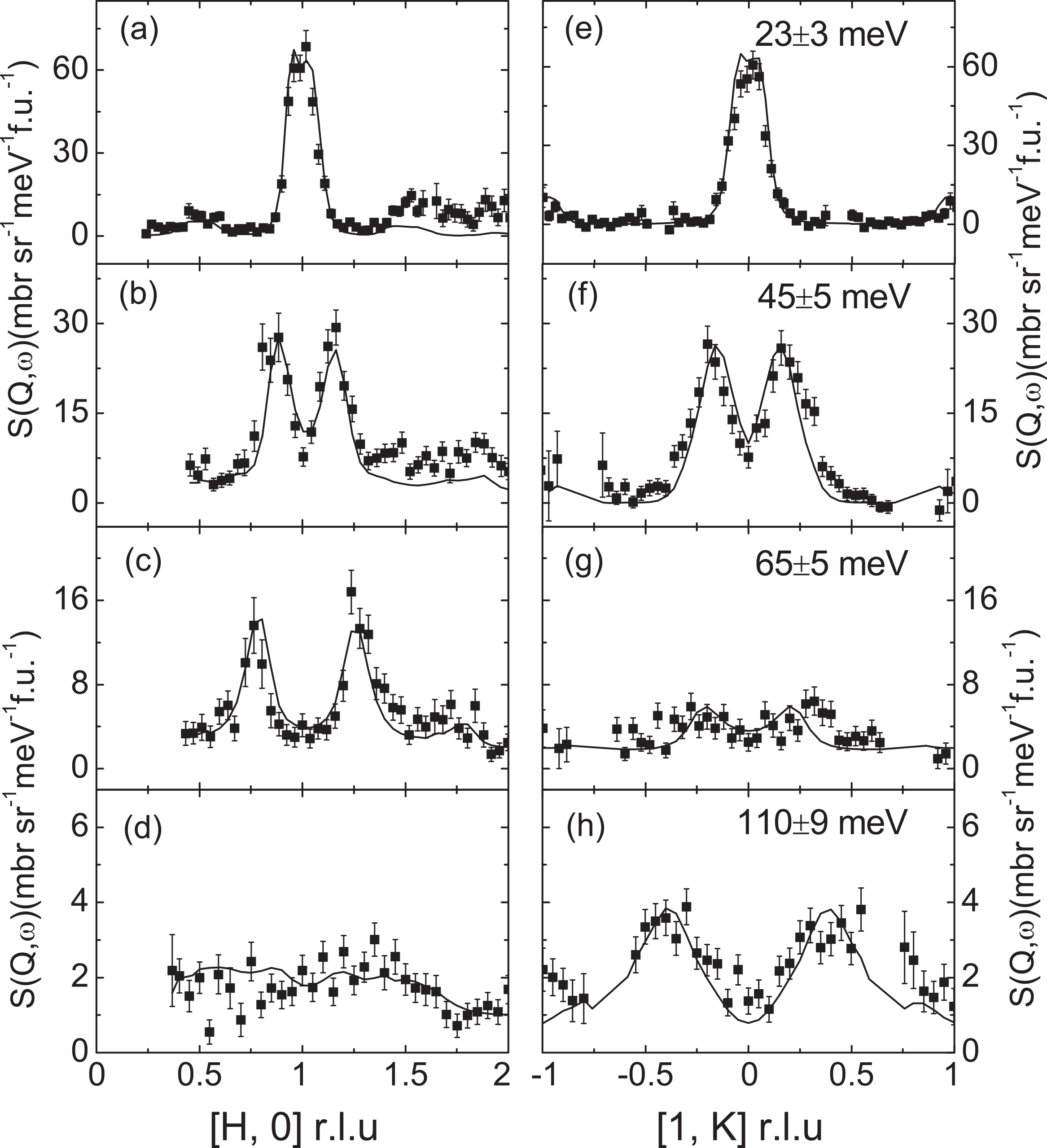}
\caption{Constant energy cuts through $Q=(1, 0)$ along the $[H, 0]$ (a-d) and $[1, K]$ (e-h) directions at energies of $E=23\pm3$ meV with $E_i=80$ meV, and $45\pm5, 65\pm5$, and $110\pm9$ meV with $E_i=250$ meV, at 8 K. The error bars indicate one sigma. The solid lines are the best fits obtained from the Tobyfit program. }
\label{fig3}
\end{figure}

To compare quantitatively the experimental data with the model, we plot cuts along the $[H, 0]$ and $[1, K]$ directions for a wide range of energies in Fig. \ref{fig3}. The solid lines are the results of the best fits with the parameters discussed above. The fits are in good agreement with the experimental data at all energies. The small discrepancy near $Q=(2, 0)$ is due to an acoustic phonon. The weaker and flatter cut along the $[1, K]$ direction at $E=65\pm5$ meV in Fig. \ref{fig3} (g) and the cut along the $[H, 0]$ direction at $E=110\pm9$ meV in Fig. \ref{fig3} (d) are consistent with the splitting of the first branch along the $[H, 0]$ direction and the 90$^\circ$ rotation of the second branch.

\begin{figure}[t]
\includegraphics[scale=0.40]{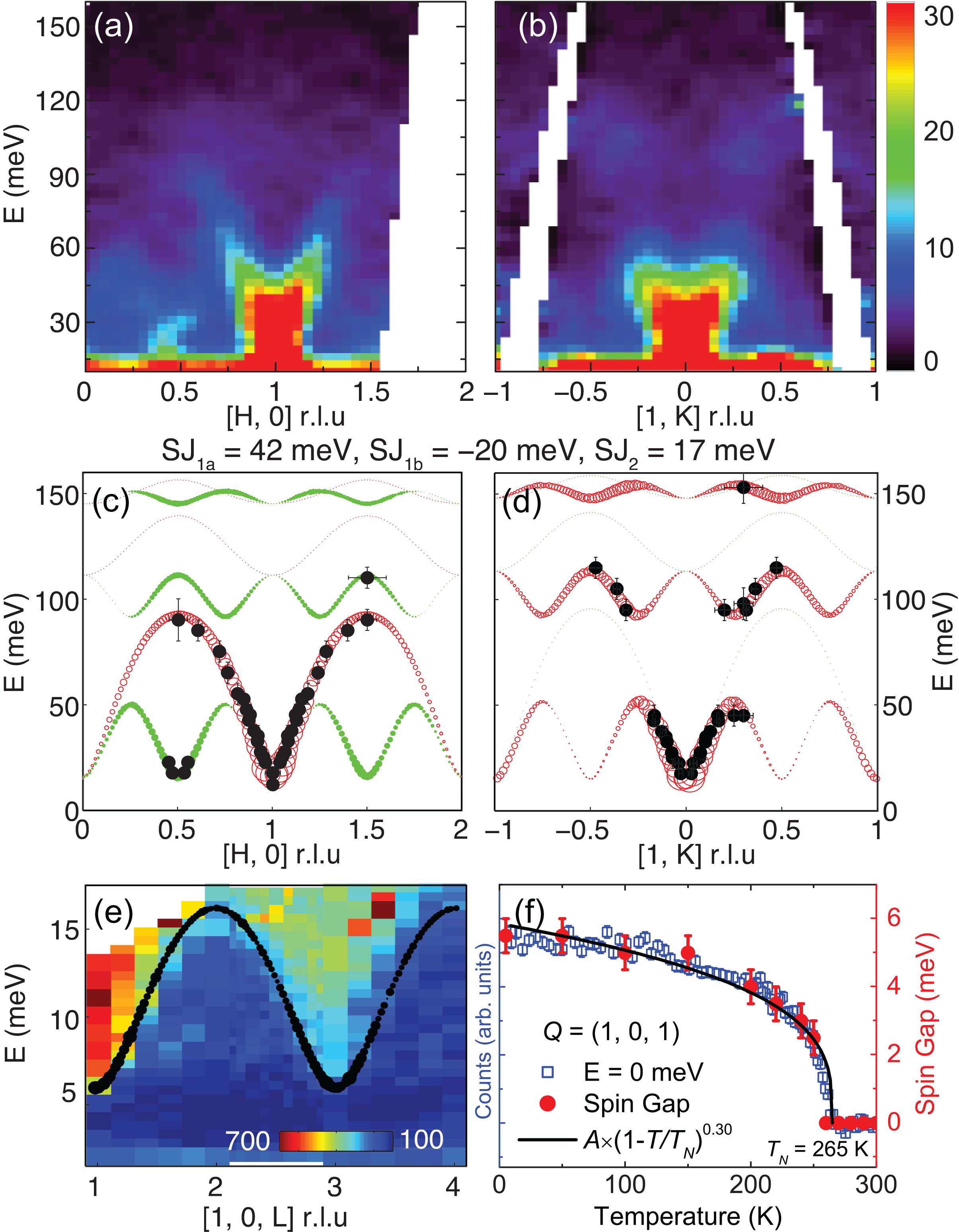}
\caption{(color online).  (a) Spin excitations along the $[H, 0]$ direction, averaging over $K=\pm0.2$ r.l.u and (b) along the $[1, K]$ direction, averaging over $H=1\pm0.2$ r.l.u with $E_i=250$ meV at 8 K. (c, d) The dispersion extracted from experimental data and simulations with the best fit parameters. The red circles are from the first twin, and the green circles are from the second twin[see supplementary information]. The intensity of the simulations is proportional to $\chi^{\prime\prime}(\bf{q}, \omega)\times\mathrm{\sqrt{E}}$. (e) The $L$-modulation of the low energy spin excitations at 2 K and simulations with $SJ_c=0.29, SJ_s=0.09$ meV and the intensity proportional to $\chi^{\prime\prime}(\bf{q}, \omega)$. (f) The temperature evolution of the spin gap measured at $Q=(1, 0, 1)$. The solid line is the result of a fit to the magnetic order parameter (the blue squares) with $A(1-T/T_N)^{\beta}$, where $A$ is a scaler, $T_N=265$ K, and $\beta=0.30$.}
\label{fig4}
\end{figure}

Fig. \ref{fig4} (a) and Fig. \ref{fig4} (b) show the dispersion relations along the $[H, 0]$ and $[1, K]$ directions with $E_i=250$ meV at 8 K, respectively. The spin excitations from the second twin at $Q=(0.5, 0), E=25$ meV in Fig. \ref{fig4} (a) and the second branch of spin excitations at energies between 90 and 120 meV in Fig. \ref{fig4} (b) can be observed. The dispersion of the spin excitations extracted from extensive constant energy cuts and $Q$ cuts, together with the results of simulations with the best fit parameters are plotted in Fig. \ref{fig4} (c) and Fig. \ref{fig4} (d). Three branches of spin excitations can be seen. We tried to fit the dispersions in Rb$_{0.8}$Fe$_{1.5}$S$_{2}$ with the parameters obtained for K$_{0.81}$Fe$_{1.58}$Se$_2$\cite{jun2}. The dispersions of the first branch along the $[H, 0]$ and $[1, K]$ directions were matched very well, but the second branch along the $[1, K]$ direction deviated from the experimental data[see supplementary information]. 

In order to determine the exchange coupling between layers, $J_c$, and the single-ion anisotropy term, $J_s$, we measured the $L$-modulation of the low energy spin excitations at 2 K. The measurements show that a gap in the spin excitations opens up below $\Delta=6$ meV and that $J_c$ only affects the spin excitation spectrum below 15 meV [Fig. \ref{fig4} (e)]. By fitting the $L$-modulated spin excitation spectrum, we determined $SJ_c=0.29\pm0.05$ and $SJ_s=0.09\pm0.02$ meV. The temperature dependence of the spin gap was also studied and is presented in Fig. \ref{fig4} (f). The spin gap remained sharp right up to the phase transition. The scaled magnetic order parameter is plotted along with the temperature dependent spin gap. The evolution of the spin gap with temperature follows the trend of the AF order, in agreement with the behavior observed in K$_2$NiF$_4$, a quasi-two-dimensional (2D) Heisenberg AF insulator\cite{bob}.

To unveil the spin state in the stripe AF order of Rb$_{0.8}$Fe$_{1.5}$S$_{2}$, we examined the sum rule of the magnetic neutron scattering. One can calculate the total fluctuating moment squared $\langle\bf{m}^2\rangle$ by integrating the susceptibility  $\chi^{\prime\prime}(\bf{q}, \omega)$ over the band width of the spin excitations via
\begin{equation}
\langle\bf{m}^\mathrm{2}\rangle=\frac{\mathrm{3\hbar}}{\pi}\int_{-\infty }^{+\infty}\frac{\int \chi^{\prime\prime}(\bf{q}, \omega)\mathrm{d}\bf{q}/\int \mathrm{d}\bf{q}}{1-exp(-\hbar\omega/\mathit{k_BT})}\mathrm{d}\omega.
\label{eq4}
\end{equation}
The total moment sum rule is $M_0^2=g^2M^2+\langle\bf{m^\mathrm{2}\rangle=\mathrm{g^2S(S+1)}}$, where $g$ is the Land$\mathrm{\acute{e}}$ $g$-factor and $M$ is the static moment. Thus the spin $S$ can be extracted\cite{miaoyin, lester, coldea}. 

The averaged dynamic susceptibility in a Brillouin zone $\chi^{\prime\prime}(\omega)=\int \chi^{\prime\prime}(\bf{q}, \omega)\mathrm{d}\bf{q}/\int \mathrm{d}\bf{q}$ is plotted in Fig. \ref{fig1} (c). The spin fluctuations in Rb$_{0.8}$Fe$_{1.5}$S$_{2}$ are obviously stronger than those in BaFe$_2$As$_2$. Integrating the dynamic susceptibility through all the spin excitation band width results in $29.7\pm5.5\mu_B^2/$formula unit (f.u.), and thus $19.8\pm3.7\mu_B^2/$Fe. Taking the ordered moment $M=2.8\pm0.5 \mu_B$ into account $M_0^2=\mathrm{(2.8\pm0.5)^2+(19.8\pm3.7)} \mu_B^2$, the total moment squared per Fe is $27.6\pm4.2 \mu_B^2$, which assuming $g=2.0$ results in a spin $\mathrm{S}=2.2\pm0.2$, which is equal to the upper limit of 24$\mu_B^2$ and $\mathrm{S}=2$ as the Hund's rule result for the 3$d$ Fe$^{2+}$ within the error. The results reveal that to within the errors all six $3d$ electrons of Fe$^{2+}$ are associated with the local moments and in the high-spin state. A candidate spin configuration is illustrated in Fig. \ref{fig1} (d). The fact that the carriers are fully localized  in Rb$_{0.8}$Fe$_{1.5}$S$_{2}$ is consistent with our photoemission measurements on several pieces of single crystals from the same batch of nearly 100$\%$ stripe AF phase. These measurements also reveal a large charge gap below the Fermi energy, suggesting that the stripe AF phase is a Mott-like insulator with the integer spin $\mathrm{S}=2$\cite{si,caochao}, rather than a small gap band insulator\cite{zlu, jzhao, meng}. The rhombic iron vacancy order stabilized at a longer in-plane Fe-Fe distance ($2.750$ \AA) of the stripe AF order larger than that of the block AF order ($2.663$ \AA) in Rb$_{0.8}$Fe$_{1.5}$S$_{2}$ could enhance the correlation and thus promotes the localization\cite{si,caochao}.

\begin{table}[t]
\caption{The magnetic exchange couplings and spin states in the stripe AF order of iron pnictides and chalcogenides\cite{jun2,leland,russell}.}
\begin{tabular}{lccccc}
\hline \hline
Compounds               & $SJ_{1a}$   & $SJ_{1b}$  & $SJ_2$ (meV) & $S$   & $M(\mu_B)$   \\ \hline
CaFe$_2$As$_2$   & $50\pm10$      & $-6\pm5$      & $19\pm4$   & 1/2  & $0.80$ \\
BaFe$_2$As$_2$   & $59\pm2$      & $-9\pm2$      & $14\pm1$   & 1/2  & $0.87$ \\
SrFe$_2$As$_2$(L)   & $31\pm1$      & $-5\pm5$      & $22\pm1$   & $0.30$  & $0.94$ \\
SrFe$_2$As$_2$(H)   & $39\pm2$      & $-5\pm5$      & $27\pm1$   & $0.69$  & $0.94$ \\
K$_{0.85}$Fe$_{1.54}$Se$_{2}$         & $38\pm7$     & $-11\pm5$     & $19\pm2$     & $-$  & 2.8\\
Rb$_{0.8}$Fe$_{1.5}$S$_2$ & $42\pm5$      & $-20\pm2$    & $17\pm2$     & 2   & $2.8\pm0.5$    \\ \hline \hline
\end{tabular}
\label{table:t1}
\end{table}

Several theoretical methods have been successfully explored to describe the spin waves of the stripe AF order: a combination of density functional theory (DFT) and dynamic mean field theory (DMFT)\cite{zhiping,chenglin}; a Heisenberg model with the anisotropic in-plane exchange couplings $J_{1a}(>0)$, $J_{1b}(<0)$, and $J_2$\cite{junnaturephys,leland,russell,jun2}; and a Heisenberg model with $J_1$, $J_2$ and a large biquadratic coupling $K$\cite{aleksander, daniel, rongyu}. The origins of the anisotropy in the $J_{1a}-J_{1b}-J_2$ model and the large biquadratic coupling in the $J_1-J_2-K$ model are still under debate. Long range nematic order is clearly not required in $X$Fe$_2$As$_2$, since the spin excitation spectrum at least at higher energies is little changed at temperatures well above the tetragonal-orthorhombic structure transition\cite{chenfang, inosov, russell, leland}. The spin waves of Rb$_{0.8}$Fe$_{1.5}$S$_{2}$ could be described by either model. In particular the rhombic iron vacancy order which has already broken the $C_4$ symmetry forms at a temperature higher than 718 K\cite{meng}. The anisotropic $J_{1a}$ and $J_{1b}$ in Rb$_{0.8}$Fe$_{1.5}$S$_{2}$ could originate from the structural orthorhombicity and the possible orbital ordering\cite{phillips}. For the $J_1-J_2-K$ model, the exchange couplings are estimated to be $J_1S=(J_{1a}+J_{1b})S/2=11\pm3$, $J_2S=17\pm2$, and $KS=(J_{1a}-J_{1b})S/4=15.5\pm1.4$ meV\cite{aleksander}.  The biquadratic term could be enhanced by the dynamic fluctuations in the chalcogen height. Distinguishing the two models microscopically is beyond the scope of this work.

We list in Table \ref{table:t1} the fitted magnetic exchange couplings and measured Fe spin values in a number of stripe phase Fe arsenides and chalcogenides.  The Fermi surfaces in these materials vary significantly as do, concomitantly, the conductivity, the ordered moments and the effective spin values.  In spite of this, the exchange couplings measured in units of $SJ$ are remarkably universal.  This result is both striking and mysterious.  It remains to be seen how this relates to the superconductivity in the doped materials.

In summary, we have studied the spin waves of the pure stripe AF order in Rb$_{0.8}$Fe$_{1.5}$S$_{2}$ over a wide range in reciprocal space and energy. Our inelastic neutron scattering data reveal that even though the stripe AF order has strikingly similar $SJ$ with all the other iron pinctides and chalcogenides, it is almost an ideal $\mathrm{S=2}$ Heisenberg antiferromagnet with fully localized moments inducing Mott insulator behavior. 

We thank Qimiao Si and Yao Shen for useful discussions. This work was supported by the Director, Office of Science, Office of Basic Energy Sciences, U.S. Department of Energy, under Contract No. DE-AC02-05CH11231 and the Office of Basic Energy Sciences U.S. DOE Grant No. DE-AC03-76SF008. We also acknowledge support from NBRPC-2012CB821400 and NSFC-11275279.



\begin{thebibliography}{99}
\bibitem{kastner} M. A. Kastner, R. J. Birgeneau, G. Shirane, and Y. Endoh, Rev. Mod. Phys. {\bf 70}, 897 (1998).
\bibitem{birgeneau} R. J. Birgeneau, C. Stock, J. M. Tranquada, and K. Yamada, J. Phys. Soc. Jpn. 75, 111003 (2006).
\bibitem{davis} D. Johnston, Adv. Phys. {\bf 59}, 803 (2010).
\bibitem{pengcheng} P. C. Dai, J. P. Hu, and E. Dagotto, Nat. Phys. {\bf 8}, 709 (2012).
\bibitem{hayden} S. M. Hayden {\it et al.}, Phys. Rev. Lett. {\bf 76}, 1344 (1996).
\bibitem{dongjing} J. Dong {\it et al.}, Europhys. Lett. {\bf 83}, 27006 (2008).
\bibitem{qimiao08} Q. Si, and E. Abrahams, Phys. Rev. Lett. {\bf 101}, 076401 (2008).
\bibitem{frank} F. Kr$\mathrm{\ddot{u}}$ger, S. Kumar, J. Zaanen, and J. van den Brink, Phys. Rev. B {\bf 79}, 054504 (2009).
\bibitem{haule} K. Haule, and G. Kotliar, New J. Phys. {\bf 11}, 025021 (2009).
\bibitem{daniel} D. Stanek, O. P. Sushkov, and G. S. Uhrig, Phys. Rev. B {\bf 84}, 064505 (2011).
\bibitem{igor} I. A. Zaliznyak {\it et al.}, Phys. Rev. Lett. {\bf 107}, 216403 (2011).
\bibitem{mengshu} M. S. Liu {\it et al.}, Nat. Phys. \textbf{8}, 376 (2012).
\bibitem{leland} L. W. Harriger {\it et al.}, Phys. Rev. B {\bf 84}, 054544 (2011); Phys. Rev. B {\bf 86}, 140403(R) (2012).
\bibitem{yjkim} H. Gretarsson {\it et al.}, Phys. Rev. Lett. {\bf 110}, 047003 (2013).
\bibitem{junnaturephys} J. Zhao {\it et al.}, Nat. Phys. \textbf{5}, 555 (2009).
\bibitem{mcqueeney2} S. O. Diallo {\it et al.}, Phys. Rev. Lett. {\bf 102}, 187206 (2009).
\bibitem{russell} R. A. Ewings {\it et al.}, Phys. Rev. B {\bf 78}, 220501(R) (2008); Phys. Rev. B {\bf 83}, 214519 (2011).
\bibitem{lei} H. C. Lei {\it et al.}, Phys. Rev. Lett. {\bf 107}, 137002 (2011).
\bibitem{jgguo} J. G. Guo {\it et al.}, Phys. Rev. B {\bf 85}, 054507 (2012).
\bibitem{jzhao} J. Zhao {\it et al.}, Phys. Rev. Lett. {\bf 109}, 267003 (2012).
\bibitem{meng} M. Wang {\it et al.}, Phys. Rev. B {\bf 90}, 125148 (2014).
\bibitem{fchen} F. Chen {\it et al.}, Phys. Rev. X {\bf 1}, 021020 (2011).
\bibitem{si} R. Yu {\it et al.}, Phys. Rev. Lett. {\bf 106}, 186401 (2011); Phys. Rev. Lett. {\bf 110}, 146402 (2013).
\bibitem{miaoyin} M. Y. Wang {\it et al.}, Nat. Commun. {\bf 2}, 580 (2011).
\bibitem{bt7} J. W. Lynn {\it et al.}, J. Research NIST {\bf 117}, 61 (2012).
\bibitem{daoxin} D. X. Yao, and E. W. Carlson, Front. Phys. China {\bf 5}, 166 (2010).
\bibitem{toby} T. G. Perring, {\it et al.}, http://tobyfit.isis.rl.ac.uk.
\bibitem{jun2} J. Zhao {\it et al.}, Phys. Rev. Lett. {\bf 112}, 177002 (2014).
\bibitem{bob} R. J. Birgeneau, J. Skalyo, and G. Shirane, Phys. Rev. B {\bf 3}, 1736 (1971).
\bibitem{lester} C. Lester {\it et al.}, Phys. Rev. B {\bf 81}, 064505 (2010).
\bibitem{coldea} J. Lorenzana, G. Seibold, and R. Coldea, Phys. Rev. B {\bf 72}, 224511 (2005).
\bibitem{caochao} C. Cao, and J. H. Dai, Phys. Rev. B {\bf 83}, 193104 (2011).
\bibitem{zlu} X. W. Yan, M. Gao, Z. Y. Lu, and T. Xiang, Phys. Rev. Lett. {\bf 106}, 087005 (2011).
\bibitem{zhiping} Z. P. Yin, K. Haule, and G. Kotliar, Nat. Phys. {\bf 10}, 845 (2014).
\bibitem{chenglin} C. Zhang {\it et al.}, Phys. Rev. Lett. {\bf 112}, 217202 (2014).
\bibitem{aleksander} A. L. Wysocki, L. D. Belashchenko, and V. P. Antropov, Nat. Phys. {\bf 7}, 485 (2011).
\bibitem{rongyu} R. Yu, Z. Wang, P. Goswami, A. H. Nevidomskyy, Q. Si, and E. Abrahams, Phys. Rev. B {\bf 86}, 085148 (2012).
\bibitem{chenfang} C. Fang, H. Yao, W. F. Tsai, J. P. Hu, and S. A. Kivelson, Phys. Rev. B {\bf 77}, 224509 (2008).
\bibitem{inosov} J. T. Park {\it et al.}, Phys. Rev. B {\bf 82}, 134503 (2010).
\bibitem{phillips} W. Lv, F. Kruger, and P. Phillips, Phys. Rev. B {\bf 82}, 045125 (2010).
\end{thebibliography}

Supplementary: Spin Waves and Spatially Anisotropic Electron Interactions in the $\mathrm{S=2}$ Stripe Antiferromagnet Rb$_{0.8}$Fe$_{1.5}$S$_2$

Meng Wang {\it et al.}

\begin{figure}[h]
\renewcommand\thefigure{S1}
\includegraphics[scale=0.35]{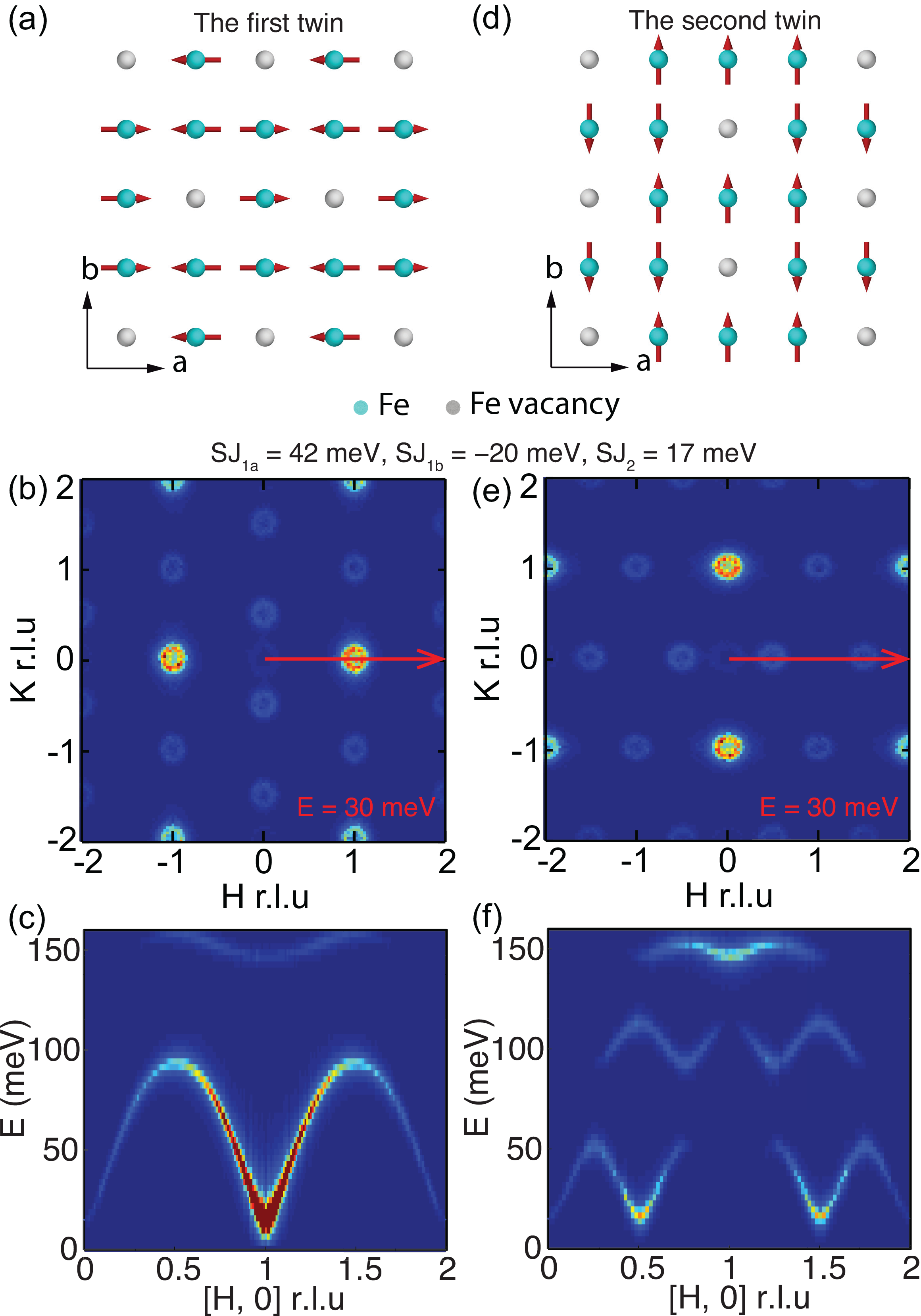}
\caption{(color online).  (a) Schematic of the iron layer with the rhombic iron vacancy order of the first twin, and (b) a simulation of the corresponding spin waves at $E=30$ meV in the $[H, K]$ plane and (c) a simulation of the dispersion along the $[H, 0]$ direction with the best fit parameters. The damping has been fixed at 3 meV for good viewability. The red arrow in (b) indicates the $Q$ in (c).  (d) Schematic of the iron layer of the second twin. (e, f) The same plots with (b) and (c) for the second twin, respectively.}
\end{figure}

\begin{figure}[b]
\renewcommand\thefigure{S2}
\includegraphics[scale=0.35]{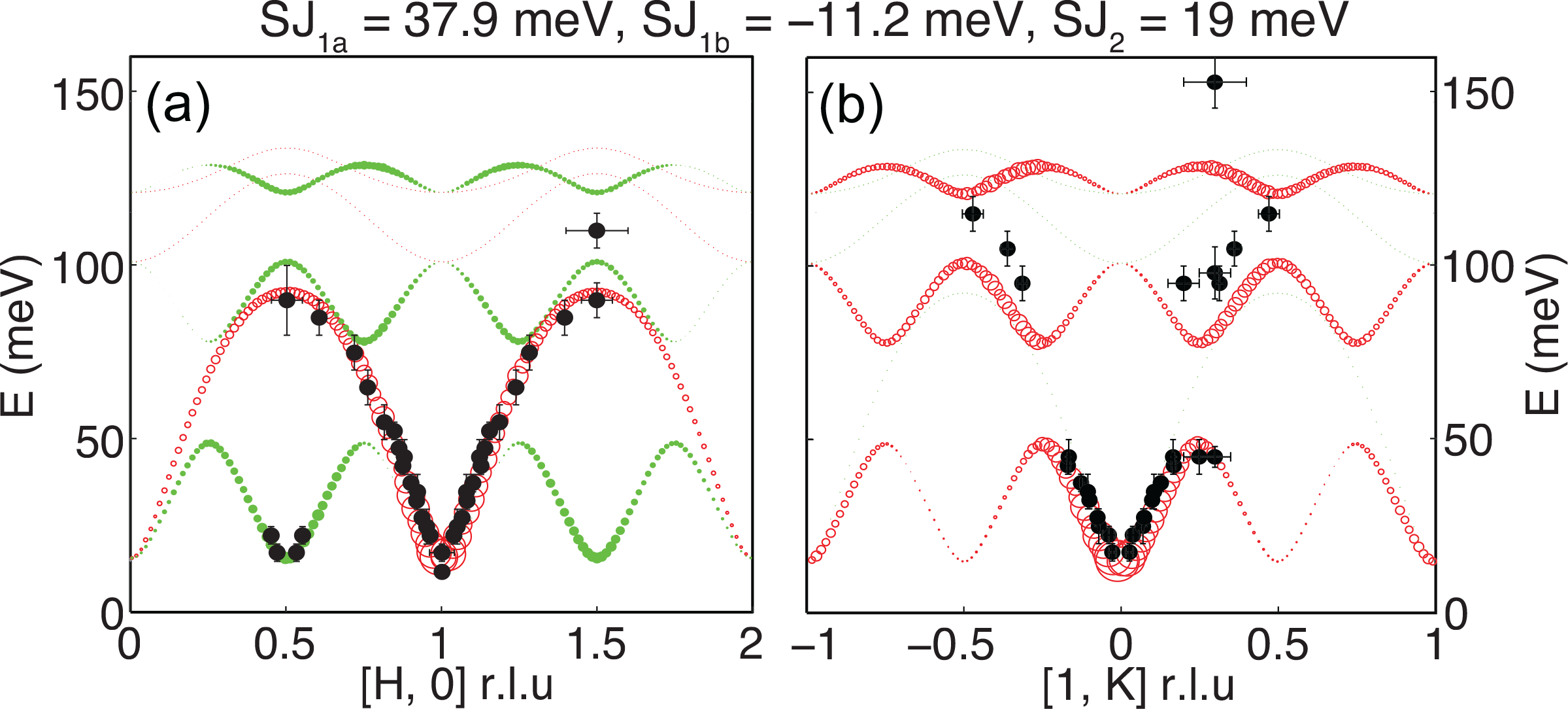}
\caption{ (color online). The dispersion extracted from the experimental data of Rb$_{0.8}$Fe$_{1.5}$S$_{2}$ and the simulations with the parameters obtained from K$_{0.81}$Fe$_{1.58}$Se$_2$[28] along the (a) $[H, 0]$ and (b) $[1, K]$ directions. The intensity of the simulations is proportional to $\chi^{\prime\prime}(\bf{q}, \omega)\times\mathrm{\sqrt{E}}$.}
\end{figure}

\end{document}